\documentclass[twocolumn,english,showpacs,floatfix]{revtex4}
\usepackage[dvips]{graphicx}
\usepackage{longtable}
\usepackage{amsmath,amssymb}
\usepackage{dcolumn}

\usepackage{babel}
\usepackage[latin1]{inputenc}

\begin{document}

\title{Stability and thermodynamics of brane black holes}

\author{E. Abdalla}
\email{eabdalla@fma.if.usp.br}
\author{B. Cuadros-Melgar}
\email{bertha@fma.if.usp.br}
\author{A. B. Pavan}
\email{alan@fma.if.usp.br}
\affiliation{Instituto de F\'{\i}sica, Universidade de S\~{a}o Paulo \\
C.P. 66318, 05315-970, S\~{a}o Paulo-SP, Brazil}

\author{C. Molina}
\email{cmolina@usp.br}
\affiliation{Escola de Artes, Ci\^{e}ncias e Humanidades, Universidade de
  S\~{a}o Paulo\\ Av. Arlindo Bettio 1000, CEP 03828-000, S\~{a}o
  Paulo-SP, Brazil}


\begin{abstract}
We consider scalar and axial gravitational perturbations of
black hole solutions in brane world scenarios. We show that perturbation
dynamics is surprisingly similar to the Schwarzschild case with strong
indications that the models are stable. Quasinormal modes and
late-time tails are discussed. We also study the thermodynamics
of these scenarios verifying the universality of Bekenstein's entropy
bound as well as the applicability of 't Hooft's brickwall method.
\end{abstract}

\maketitle

\section{Introduction}

The extra dimensions idea had its origin in the seminal works by
Kaluza and Klein in the 20's \cite{KK} and gained momentum in the
context of string theory in the last decades \cite{polchinski}.
Recent developments on higher--dimensional gravity resulted in a
number of interesting theoretical ideas such as the brane world
concept. The essence of this string inspired model is that
Standard Model fields are confined to a three dimensional
hypersurface, the brane, while gravity propagates in the full
spacetime, the bulk. The simplest models in this context
(abbreviated here as RSI and RSII), proposed by Randall and Sundrum
\cite{rs}, describe our world as a domain wall embedded in a
$Z_{2}$-symmetric five dimensional anti--de Sitter (AdS) spacetime.
The RSI model proposes a mechanism to solve the hierarchy problem
by a small extra dimension, while the RSII model considers an infinite
extra dimension with a warp factor which ensures the localization
of gravity on our brane.

Black holes are important sources of gravitational waves that are
expected to be detected by the current and upcoming generation of
experiments. This will open up a new window for testing
modifications of general relativity. The simplest case of
gravitational collapse in the standard four dimensional world is
described by the 4-dimensional Schwarzschild metric. In the
5-dimensional scenario it would be natural to ask whether matter
confined on the brane after undergoing gravitational collapse can still
be described by a Schwarzschild-type metric. The most natural
generalization in the RSII model corresponds to a black string
infinite in the fifth dimension, whose induced metric on the brane
is purely Schwarzschild \cite{chr}. However, although the
curvature scalars are everywhere finite, the Kretschmann scalar
diverges at the AdS horizon at infinity, which turns the above
solution into a physically unsuitable object.

It has been argued that there exists a localized black cigar
solution with a finite extension along the extra dimension due to
a Gregory-Laflamme \cite{glf} type of instability near the AdS
horizon. A class of such a solution has been found by Casadio
\textit{et al.} \cite{cfm1,cfm2} using the projected Einstein
equations on the brane derived by Shiromizu \textit{et al.}
\cite{sasaki}. It has the desired ``pancake'' horizon structure
ensuring a non-singular behavior in the curvature and Kretschmann
scalars at least until the order of the multipole expansion
considered there. In fact, this solution belongs to a class of
black hole solutions found later by Bronnikov \textit{et al.}
\cite{bron}, who also classified several possible brane
black holes obtained from the Shiromizu \textit{et al.} projected
equations \cite{sasaki} in two families according to the horizon
order. For such spacetimes only horizons of order 1 or 2 are
possible, but not higher than that.

In this paper we are interested in the study of black holes from the
point of view of a brane observer, as ourselves. We analyze some
characteristics of Bronnikov \textit{et al.} solutions. Firstly,  
some aspects of the thermodynamics are
studied. Black Hole Thermodynamics was constructed when Bekenstein
proposed the proportionality law between the entropy and the horizon area
\cite{bekenstein}. The discovery of Hawking radiation validated this
proposal and established the proportionality factor $1/4$ in a
precise way \cite{hawking} leading to the well-known
Bekenstein-Hawking formula,
\begin{equation}\label{bekhaw}
S_{BH} = \frac{\textrm{Area}}{4G} \, .
\end{equation}

One way to compute the entropy based on a semi-classical
description of a scalar field was proposed by 't Hooft
\cite{thooft} and it is known as the brick wall method, which was
frequently used later in several contexts \cite{alej}. When
applying this method to a Schwarzschild black hole, 't Hooft found
that the entropy was proportional to the area, as expected, but
additionally it had a $\alpha^{-2}$ correction, $\alpha$ being the
proper distance from the horizon to the wall. This term was later
interpreted as a one-loop correction to the Bekenstein-Hawking
formula, since it can be absorbed as a renormalization of the
gravitational coupling constant $G$ \cite{susuglu}.

Another interesting feature of black hole thermodynamics is the
existence of an upper bound on the entropy of any neutral system of
energy $E$ and maximal radius $R$ in the form $S\leq 2\pi ER$,
proposed by Bekenstein \cite{bekenstein2}. This bound becomes
necessary in order to enforce the generalized second law of
thermodynamics (GSL).

Besides thermodynamical results, we also consider the response of
a brane black hole perturbation which should represent some damped oscillating
signal. It can be decomposed with Laplace
transformation techniques into a set of so-called quasinormal
modes (QNM).  The QNMs of black holes are important because they
dominate in the intermediate late-time decay of a perturbation
and do not depend upon the way they were excited. They depend only
on the parameters of a black hole and are, therefore, the
``footprints'' of this structure.  The time-independent problem
for perturbations of a brane black hole turns out to be quite
similar to that for a black hole: one has to find the solutions of
the wave-like equations satisfying the appropriate boundary
conditions, which we shall further discuss in detail.

The paper is organized as follows, in Sec. II the brane black
holes are presented. Sec. III discusses the thermodynamical
properties of the solutions thus considered. Sec. IV treats the
question of perturbation and stability of these objects. Wave-like
equations for the perturbations are derived. In Sec. V an analysis
of the quasinormal modes and late-time behavior is developed. In
Sec. VI we summarize our results, and some final comments are made.

\section{Brane black hole solutions}

The vacuum Einstein equations in $5$ dimensions, when projected on
a 4-dimensional spacetime and after introducing gaussian
normal coordinates ($x^{\mu}$, with $\mu=0 \ldots 3$, and $z$), lead to the
gravitational equation on the 3-brane given by \cite{sasaki}
\begin{equation}\label{h-cons}
R_{\mu\nu} ^{(4)} = \Lambda_4 g_{\mu\nu} ^{(4)} - E_{\mu\nu} \,
\end{equation}
where $\Lambda_4$ is the brane cosmological constant, and $E_{\mu\nu}$
is proportional to the (traceless) projection on the brane of the
$5$-dimensional Weyl tensor.

The only combination of the Einstein equations in a brane world that
can be written unambiguously without specifying $E_{\mu\nu}$ is their
trace \cite{cfm1,cfm2,bron},
\begin{equation}\label{trace}
R^{(4)}=4\Lambda_4 \, .
\end{equation}
It is clear that this equation, also known as the hamiltonian
constraint in the ADM decomposition of the metric, is a weaker
restriction than the purely 
4-dimensional equation $R_{\mu\nu}=0$, which, in fact, is equivalent
to Eq. (\ref{h-cons}), provided that we know the structure of $E_{\mu\nu}$.

In order to obtain four dimensional solutions of Eq. (\ref{trace}), 
we choose the spherically symmetric form of the 4-dimensional metric
given by 
\begin{equation}\label{genmet}
ds^2 = -A(r) dt^2 + \frac{dr^2}{B(r)} + r^2 (d\theta^2 + \sin^{2} \theta \,
d\phi^2) \quad .
\end{equation}
We relax the condition $A(r)=B(r)$, which is accidentally verified
in four dimensions but, in fact, there is no reason for it to
continue to be valid in this scenario. In this spirit, black hole
and wormhole solutions \cite{cfm1,bron,bronworm,gregory} as well as star
solutions on the brane \cite{germar} have been obtained in the
last years. We should mention that even without relaxing this
condition, previous solutions have also been found
\cite{others,bhsym}.

In this context the hamiltonian constraint can be written explicitly
in terms of $A$ and $B$ as \cite{cfm1,bron}
\begin{gather}
B \left[ \frac{A''}{A} - \frac{1}{2} \left(
  \frac{A'}{A}\right)^2 + \frac{1}{2} \frac{A'}{A}
  \frac{B'}{B} + \frac{2}{r} \left( \frac{B'}{B} +
  \frac{A'}{A} \right) \right] \nonumber \\
-\frac{2}{r^{2}} (1 - B) =  4\Lambda_4 \, ,
\label{hamcons}
\end{gather}
with prime (') denoting differentiation with respect to $r$.

We will center our attention in the black hole type solutions which
can be obtained by one of the following algorithms \emph{BH1} and \emph{BH2}.
These are subclasses of the corresponding algorithms in \cite{bron}
(where the parameter $s$ in this reference is set to 1). \\  
\emph{First Algorithm (BH1)}: Specify a function $A(r)$, positive and
  analytical in a neighborhood of the event horizon
  ${\cal R}[r]$, in such a way that $4A + rA'>0$ in ${\cal
  R}[r]$, and $A \approx (r-r_h)$  as $r\rightarrow
  r_h$. Then $B(r)$ is given by the general solution of
  (\ref{hamcons}) with vanishing brane cosmological constant,
\begin{gather}
B(r) = \frac{2A e^{3\Gamma}}{r(4A+rA')^2} \nonumber \\
\times \left[ \int_{r_h}^{r} (4A + rA')(2-r^2R) e^{-3\Gamma} dr + C
\right] \, , \label{b1}
\end{gather}
where $C$ is an integration constant and
\begin{equation}
\Gamma(r) = \int \frac{A'}{4A+rA'} \, dr \, .
\label{b}
\end{equation}
For $C\geq 0$ we have a black hole metric with a horizon at
$r=r_h$, which is simple if $C>0$ and of the order $2+p$ if both
$C=0$, and $Q$ has the behavior,
\begin{equation}\label{q}
Q(r) = 2-r^2 R \approx (r-r_h)^p, \; \hbox{near} \; r=r_h \, , \;
p\in\textbf{N} \, .
\end{equation}
\emph{Second Algorithm (BH2)}:  Specify a function $A(r)$, positive and
  analytical  in a neighborhood of the event horizon
  ${\cal R}[r]$, in such a way that $4A + rA'>0$ in ${\cal
  R}[r]$, and $A\approx (r-r_h)^{2}$ as $r\rightarrow r_h$. Then
  $B(r)$ is again given by (\ref{b1}). The black hole metric appears
  when $C=0$ with a horizon at $r=r_h$ of the order $2+p$ if $Q(r)$
  behaves according to (\ref{q}).

Both algorithms lead to double horizons in the case $C=0$, if $Q(r_h)>0$. A case in point of the first algorithm is the solution
with the metric element $A(r)$ having the usual form of a
Schwarzschild black hole found by Casadio, Fabbri, and Mazzacurati
(CFM solution) \cite{cfm1,cfm2} given by
\begin{eqnarray}\label{cfm}
A(r)&=&1-\frac{2M}{r}, \nonumber \\
B(r)&=&\frac{(1-\frac{2M}{r}) (1-\frac{M\gamma}{2r})} {1-\frac{3M}{2r}}\, ,
\end{eqnarray}
where $\gamma$ is an integration constant. The event
horizon is localized at $ r_{h} =2M$ and the singularity at $r=3M/2$
instead of $r=0$. Notice that the Schwarzschild solution is recovered
with $\gamma=3$. In this work we are restricted to the case when
$\gamma < 4$.

Another interesting example of this algorithm is the metric with zero
Schwarzschild mass \cite{bron} given by
\begin{eqnarray}
A(r) &=& 1-\frac{h^2}{r^2} \, , \quad h>0 \, , \nonumber \\
B(r) &=& \left( 1 - \frac{h^2}{r^2} \right) \left( 1+
\frac{C-h}{\sqrt{2r^2-h^2}} \right) \, ,
\label{zeroSch}
\end{eqnarray}
whose horizon $r=h$ is simple if $C>0$ and double if $C=0$. The
singularity occurs at $r=h/\sqrt{2}$. This example shows that in
the brane world context a black hole may exist without matter and
without mass, only as a tidal effect from the bulk gravity. However,
there is a special situation when $h^2$ can be related to a
5-dimensional mass, namely, $C=h$. In this case Eq. (\ref{zeroSch}) is
the induced metric of a 5-dimensional Schwarzschild black hole, as
described in \cite{roman}, where the chosen background was ADD-type.

\section{Black hole thermodynamics}

In order to study the thermodynamical properties of the brane black
holes generated by the BH1 and BH2 algorithms, we use the following
expressions of the metric coefficients near the horizon
\begin{eqnarray}\label{nearhor}
A(r)&=& A_1 (r-r_h) + {\cal O}((r-r_h)^2) \\
B(r)&=& B_1 C (r-r_h) + B_2 (r-r_h)^2 + {\cal O}((r-r_h)^3) \, , \nonumber
\end{eqnarray}
for BH1 algorithm with $A_1, B_1, B_2 >0$ and $C$ being an integration
constant that defines the black hole family, and
\begin{eqnarray}
A(r) &=& A_2 (r-r_h)^2 + {\cal O}(r-r_h)^3 \, , \nonumber \\
B(r) &=& B_3 C + B_4 (r-r_h)^2 + B_5 (r-r_h)^3 \nonumber \\
     & & + {\cal O}((r-r_h)^4) \, ,
\label{bh2}
\end{eqnarray}
for BH2 algorithm, being $C$ the family parameter again. 

We will show here the calculation for the BH2 family, which turns out
to be more interesting, since the metric coefficients expansion
(\ref{bh2}) is different from the standard one
(\ref{nearhor}). However, we will display certain quantities for both
families wherever it is relevant.

We first consider the issue of the entropy bound.
The surface gravity at the event horizon is given by
\begin{equation}
\kappa = \begin{cases}
          \frac{1}{2} \sqrt{A_1 B_1 C} & \textrm{for BH1 family}\, , \\
          \sqrt{A_2 B_3 C} &  \textrm{for BH2 family}\, .
         \end{cases} 
\label{kappa}
\end{equation}

Let us consider an object with rest mass $m$ and proper radius $R$
descending into a BH2 black hole. The constants
of motion associated to $t$ and $\phi$ are \cite{qwa}
\begin{equation}\label{EJ}
E = \pi_t \, , \quad J=-\pi_\phi \, ,
\end{equation}
where
\begin{eqnarray}\label{pis}
\pi_t &=&g_{tt} \dot t \, , \nonumber \\
\pi_\phi &=& g_{\phi\phi} \dot \phi \, .
\end{eqnarray}
In addition,
\begin{equation}\label{m}
m^2 = -\pi_\mu \pi^{\mu} \, .
\end{equation}

For simplicity we just consider the equatorial motion of the object,
\textit{i.e.}, $\theta=\pi/2$. The quadratic equation for the
conserved energy $E$ of the body coming from (\ref{EJ})-(\ref{m}) is
given by
\begin{equation}\label{ee}
\alpha E^2 -2\beta E + \zeta =0 \, ,
\end{equation}
with
\begin{eqnarray}\label{abg}
\alpha&=& r^2 \, , \nonumber \\
\beta&=& 0 \, ,  \\
\zeta &=& A_2 (r-r_h)^2 (J^2+m^2 r^2) \, . \nonumber
\end{eqnarray}

The gradual approach to the black hole must stop when the proper
distance from the body's center of mass to the black hole horizon
equals $R$, the body's radius,
\begin{equation}\label{dint}
\int_{r_h}^{r_h + \delta(R)} \frac{dr}{\sqrt{B(r)}} = R \, .
\end{equation}
Integrating this equation we obtain the expression for $\delta$,
\begin{equation}\label{delta}
\delta = \begin{cases}
          \frac{1}{4} C R^2 B_1 & \textrm{for BH1 family} \, , \\
          R \sqrt{B_3 C} & \textrm{for BH2 family} \, .
         \end{cases}	  
\end{equation}

Solving (\ref{ee}) for the energy and evaluating at the point of
capture $r=r_h +\delta$ we have
\begin{equation}\label{ecap}
E_{cap} \approx \frac{\sqrt{A_2(J^2+m^2 r_h ^2)}\; \delta}{r_h}
\end{equation}
This energy is minimal for a minimal increase in the black hole
surface area, $J=0$, such that
\begin{equation}\label{emin}
E_{min} = \sqrt{A_2} \, m \,\delta \, .
\end{equation}

From the First Law of Black Hole Thermodynamics we know that
\begin{equation}\label{fl}
dM = \frac{\kappa}{2} dA_r \, ,
\end{equation}
where $A_r$ is the rationalized area ($\textrm{Area}/4\pi$), and $dM = E_{min}$ is
the change in the black hole mass due to the assimilation of the
body. Thus, using (\ref{kappa}) we obtain
\begin{equation}\label{area}
dA_r = 2mR \, .
\end{equation}

Assuming the validity of the GSL, $S_{BH}(M+dM) \geq S_{BH}(M) +S$, we
derive an upper bound to the entropy $S$ of an arbitrary system of
proper energy $E$,
\begin{equation}\label{entrop}
S \leq 2\pi E R \, .
\end{equation}
This result coincides with that obtained for the purely 4-dimensional
Schwarzschild solution, and it is also independent of the black hole
parameters \cite{bekenstein2}. \emph{It shows that the bulk does not affect
the universality of the entropy bound.}

Let us find now the quantum corrections to the classical BH
entropy. We consider a massive scalar field $\Phi$ in the background
of a BH2 black hole satisfying the massive Klein-Gordon equation,
\begin{equation}
\left( \Box - m^2 \right) \Phi = 0 \, .
\label{kg}
\end{equation}
In order to quantize
this scalar field we adopt the statistical mechanical approach
using the partition function $Z$, whose leading contribution comes
from the classical solutions of the euclidean lagrangian that
leads to the Bekenstein-Hawking formula. In order to compute the
quantum corrections due to the scalar field we use the 't Hooft's
brick wall method, which introduces an ultraviolet cutoff near the
horizon, such that
\begin{equation}\label{uv}
\Phi(r)=0 \quad \hbox{at} \quad r=r_h +\varepsilon \, ,
\end{equation}
and an infrared cutoff very far away from the horizon,
\begin{equation}\label{ir}
\Phi(r)=0 \quad \hbox{at} \quad r=L \gg M \, .
\end{equation}

Thus, using the black hole metric (\ref{genmet}) and the
\textit{Ansatz} $\, \Phi= e^{-iEt} R(r) Y_{\ell m}(\theta,\phi)$,
Eq.(\ref{kg}) turns out to be
\begin{gather}
\frac{E^2}{A} R + \sqrt{\frac{B}{A}}  \frac{1}{r^2} \partial_r
\left( r^2 \sqrt{AB}  \partial_r R \right) \nonumber \\
- \left[ \frac{\ell (\ell +1)}{r^2} + m^2 \right] R =0 \, .
\label{kg3}
\end{gather}

Using a first order WKB approximation with $R(r) \approx e^{iS(r)}$ in
(\ref{kg3}) and taking the real part of this equation we can obtain
the radial wave number $K\equiv \partial_r S$ as being,
\begin{equation}\label{K}
K = B^{-1/2} \left[ \frac{E^2}{A} - \left( \frac{\ell (\ell +1)}{r^2} + m^2
\right) \right]  ^{1/2} \, .
\end{equation}

Now we introduce the semiclassical quantization condition,
\begin{equation}\label{qs}
\pi \, n_r = \int_{r_h+\varepsilon} ^L K(r,\ell,E) \, dr \, .
\end{equation}

In order to compute the entropy of the system we first calculate the
Helmholtz free energy $F$ of a thermal bath of scalar particles with
temperature $1/\beta$,
\begin{equation}\label{Helm}
F = \frac{1}{\beta} \int d\ell (2\ell+1) \int dn_r \ln (1-e^{-\beta E}) \, .
\end{equation}
Integrating by parts, using (\ref{K}) and (\ref{qs}), and performing
the integral in $\ell$ we have
\begin{equation}\label{Helm3}
F = - \frac{2}{3\pi} \int_{r_h + \varepsilon} ^L dr \, A^{-3/2} B^{-1/2}
r^2 \int dE \, \frac{(E^2- A m^2)^{3/2}}{e^{\beta E}-1} \, .
\end{equation}

Following 't Hooft's method, the contribution of this
integral near the horizon is given by 
\begin{equation}\label{Helm5}
F \approx -\frac{2r_h ^3}{3\pi} \int_{1+\bar\varepsilon} ^{\bar L} dy
  \frac{(A_2 r_h ^2)^{-3/2} (y-1)^{-3}}{(B_3 C)^{1/2}} \int_0 ^\infty dE
  \frac{E^3}{e^{\beta E}-1} \, ,
\end{equation}
where $y=r/r_h$, $\bar \varepsilon = \varepsilon /r_h$, and $\bar L =
L/r_h$. Notice that since $A$ goes to $0$ near the horizon, the
mass term in Eq. (\ref{Helm3}) becomes negligible. 

Therefore, the leading divergent contribution to $F$ (with $\varepsilon \rightarrow
0$) is  
\begin{equation}\label{F}
F_\varepsilon= -\frac{r_h ^2 \pi^3}{45 \beta^4}
\frac{(A_2)^{-3/2}}{(B_3 C)^{1/2} \varepsilon ^2} \, .
\end{equation}

The corresponding entropy is then,
\begin{equation}\label{entro}
S_\varepsilon = \beta^2 \frac{\partial F}{\partial \beta} = \frac{4
  r_h ^2 \pi^3 (A_2)^{-3/2}}{45 (B_3 C)^{1/2} \varepsilon ^2 \beta^3}
\, .
\end{equation}
Using the value of the Hawking temperature $T_H = 1/\beta
=\kappa/2\pi$, with $\kappa$ given in (\ref{kappa}) we have
\begin{equation}\label{S1}
S_\varepsilon = \begin{cases}
                 r_h ^2 B_1 C/(360 \varepsilon) & \textrm{for
                 BH1 family} \, , \\
                 r_h ^2 B_3 C/(90\, \varepsilon ^2) &
                 \textrm{for BH2 family} \, .
                \end{cases}
\end{equation}

We can express our result in terms of the proper thickness $\alpha$
given by
\begin{equation}\label{alpha}
\alpha = \int_{r_h} ^{r_h+\varepsilon} \frac{dr}{\sqrt{B(r)}} \approx
\begin{cases}
2\sqrt\varepsilon/\sqrt{B_1 C} & \textrm{for BH1 family} \, , \\
\varepsilon /\sqrt{B_3 C} & \textrm{for BH2 family} \, .
\end{cases}
\end{equation}
Thus,
\begin{equation}\label{S2}
S_\varepsilon = \frac{r_h ^2}{90 \alpha^2}\, ,
\end{equation}
or in terms of the horizon area $\textrm{Area}=4\pi r_h ^2$,
\begin{equation}\label{S3}
S_\varepsilon = \frac{\textrm{Area}}{360\pi \alpha^2}\, ,
\end{equation}
which is the same quadratically divergent correction found by 't
Hooft \cite{thooft} for the Schwarzschild black hole and by Nandi
\textit{et al.} \cite{nandi} for the CFM brane black hole. Thus,
we see that the correction is linearly dependent on the area.

The calculation of the entropy bound and entropy quantum correction
for the BH1 black hole is similar and leads to the same
results shown in (\ref{entrop}) and (\ref{S3}).

\section{Perturbative dynamics: matter and gravitational perturbations}

For simplicity we model the matter field by a scalar field $\Phi$
confined on the brane obeying the massless \mbox{($m = 0$)} version of the
Klein-Gordon equation (\ref{kg}). \emph{We expect that  massive
fields ($m\ne0$)  should show  rather different tail behavior, but
such cases will not be treated in the present paper.}

Using the decomposition of the scalar field as
$\Psi(t,r, \theta,\phi) = R(t,r) \textrm{Y}_{\ell,m}(\theta,\phi)$ in
terms of the angular, radial, and time variables we have the equation
\begin{equation}
-\frac{\partial^{2} R_{\ell}}{\partial
 t^{2}}+\frac{\partial^{2} R_{\ell}}{\partial
 r_{\star}^{2}}=V_{sc} (r(r_\star))R_{\ell} \, ,
\label{wave_equation1}
\end{equation}
with the tortoise coordinate $r_\star$ defined as
\begin{equation}
\frac{d r_{\star} (r)}{dr} = \frac{1}{\sqrt{A(r) B(r)}} \, .
\label{rstar}
\end{equation}
The effective potential $V_{sc}$ is given by
\begin{equation}
V_{sc} = A(r) \frac{\ell (\ell + 1)}{r^{2}}
+ \frac{1}{2 r} \left[ A(r) \, B'(r) + A'(r) \, B(r) \right] \, .
\label{pot_escalar}
\end{equation}

In order to address the problem of black hole stability under
gravitational perturbations, we consider first order perturbation
of $R_{\alpha\beta} = -E_{\alpha\beta}$, where $R_{\alpha\beta}$ and
$E_{\alpha\beta}$ are the Ricci tensor and the projection of the five
dimensional Weyl tensor on the brane, respectively. In general, the
gravitational perturbations depend on the tidal perturbations, namely,
$\delta E_{\alpha\beta}$. Since the complete bulk solution is not
known, we shall use the simplifying assumption $\delta
E_{\alpha\beta}=0$. This  assumption can be justified at least in a
regime where the perturbation energy does not exceed the threshold of
the Kaluza-Klein massive modes. Analysis of gravitational shortcuts
\cite{shortcuts} also supports this simplification showing that
gravitational fields do not travel deep into the bulk. On the other
hand, since we ignore bulk back-reaction, the developed perturbative
analysis should not describe the late-time behavior of gravitational
perturbations. Within such premises we obtain the gravitational
perturbation equation
\begin{equation}\label{gravperteq}
\delta R_{\alpha\beta} = 0 \, .
\end{equation}

We will consider axial perturbations in the brane geometry, following the
treatment in \cite{chandra}. To make this section more self contained,
we will briefly describe the treatment used in \cite{chandra}. 
The metric \textit{Ansatz} with sufficient generality is 
\begin{eqnarray}
ds^2 & = & e^{2\nu} dt^2 - e^{2\psi} 
\left( d\phi^2 - \omega dt - q_{2} d\theta^{2} - q_{3} d\phi^2 \right)
\nonumber \\
& & - e^{2\mu_{2}}dr^{2} - e^{2\mu_{3}} d\theta^{2}
\label{pert-metric}
\end{eqnarray}
where we adopt here a more convenient notation,
\begin{equation}
e^{2\nu} = A(r) \, , \, e^{2\mu_{2}} = \frac{1}{B(r)} \, , \,
e^{2\mu_{3}} = r^{2}  \, , \, e^{2\psi} = r^{2} \sin^{2} \theta \, .
\end{equation}
Axial perturbations in the brane metric (\ref{genmet}) are characterized by
non-null (but first order) values for $\omega$, $q_{2}$, and $q_{3}$
in Eq. (\ref{pert-metric}). We refer to \cite{chandra} for further
details.

In order to decouple the system, it is adopted the change of variables
\begin{equation}
Q_{\alpha\beta} = q_{\alpha,\beta} - q_{\beta,\alpha} \, ,
\end{equation}
and 
\begin{equation}
Q_{\alpha\,0} = q_{\alpha,0} - \omega,_{\alpha} \, , 
\end{equation}
with $\alpha,\beta = 2,3$. We denote partial differentiation with respect to
$t$, $\theta$ and $\phi$  by ``,0'', ``,1'' and ``,2'', respectively.
The perturbations are then described by  
\begin{gather}
(e^{3\psi +\nu-\mu_{2}-\mu_{3}}Q_{23})_{,3}
e^{-3\psi+\nu-\mu_{3}+\mu_{2}}= \nonumber\\
=- (\omega_{,2}-q_{2,0})_{,0}\ ,
\label{eq_1}
\end{gather}
\begin{gather}
(e^{3\psi
  +\nu-\mu_{2}-\mu_{3}}Q_{23})_{,2}e^{-3\psi+\nu+\mu_{3}-\mu_{2}} =
  \nonumber\\
=  (\omega_{,3}-q_{3,0})_{,0}\ .
\label{eq_2}
\end{gather}

Setting $Q(t,r,\theta)= \exp (3\psi +\nu-\mu_{2}-\mu_{3} )Q_{23}$,
Eqs.(\ref{eq_1}) and (\ref{eq_2}) can be combined as
\begin{gather}
r^4\sqrt{\frac{B(r)}{A(r)}}\frac{\partial}{\partial
  r}\left[\frac{1}{r^2}\sqrt{A(r)B(r)} \, \frac{\partial
  Q}{\partial r}\right]-r^2\sqrt{\frac{B(r)}{A(r)}}\frac{\partial^2
  Q}{\partial t^2} = \nonumber\\
= -\sin^3 \theta \frac{\partial}{\partial
  \theta}\left(\frac{1}{\sin^3 \theta}\frac{\partial Q}{\partial
  \theta}\right) \quad .
\label{almost_eq}
\end{gather}
We further separate variables and write Eq.(\ref{almost_eq}) in
the form of a Schr\"{o}dinger-type wave equation by introducing
\mbox{$Q(t,r,\theta)= r Z_{\ell}(t,r) C^{-3/2}_{\ell+2}(\theta)$}, 
and $r=r(r_{\star})$, where $C^{-3/2}_{\ell+2}(\theta)$ is the
Gegenbauer function. Thus, the axial gravitational perturbations
are given by an equation of motion with the form given in
(\ref{wave_equation1}) with the effective potential
\begin{gather}
 V_{grav}(r) = A(r) \frac{(\ell + 2)(\ell - 1)}{r^{2}}
 + \frac{2 A(r) B(r)}{r^{2}}\nonumber \\
  - \frac{1}{2 r} \left[  A(r) \, B'(r) + A'(r) \, B(r) \right] \, .
\label{pot_axial}
\end{gather}

\section{Specific models}

\subsection{Overview of the results}

The equations of motion for the scalar and axial gravitational
perturbations give us a tool to analyze the dynamics and stability
of the black hole solutions in both the CFM and ``zero mass'' black
hole backgrounds. Of particular interest are the quasinormal
modes. They  are defined as the solutions  of
Eq. (\ref{wave_equation1}) which satisfy both boundary conditions
that require purely out-going waves at (brane) spatial infinity
and purely in-going waves at the event horizon,
\begin{equation}
\lim_{x \rightarrow \mp \infty} \Psi \, e^{\pm i \omega
  x} = \textrm{constant} \, ,
\end{equation}
with $\Psi=R_{\ell}$ and  $Z_{\ell}$ for the scalar and gravitational
perturbations, respectively.

In order to analyze quasinormal mode phase and late-time behavior
of the perturbations, we apply a numerical characteristic
integration scheme based in the light-cone variables $u = t -
r_\star$ and $v = t + r_\star$ used, for example, in
\cite{Pullin,molina,abdetal}. In addition, to check some results obtained
in ``time--dependent'' approach we employ the semi-analytical
WKB-type method developed in \cite{wkb1} and improved in
\cite{wkb2}. Both approaches show good agreement for the
fundamental overtone which is the dominating contribution in the
signal for intermediate late-time.

At a qualitative level we have observed the usual picture in the
perturbative dynamics for all fields and geometries considered here. After
the initial transient regime, it follows the quasinormal mode phase
and finally a power-law tail.  In contrast to the 5-dimensional model
in \cite{marteens}, in the present context we do not observe
Kaluza-Klein massive modes in the late-time behavior of the
perturbations. This is actually expected, since our treatment for
gravitational perturbations neglects the back-reaction from the
bulk. Still, as discussed in section IV, our results should model the
quasinormal regime. 

A necessary condition for the stability of the geometries we have
considered is that these backgrounds must be stable under the
perturbations modelled by the effective potentials
(\ref{pot_escalar}) and (\ref{pot_axial}).

If the effective potential ($V$) is positive definite, the
differential operator
\begin{equation}
\mathcal{D} = -\frac{\partial^{2}}{\partial r_{\star}^{2}} + V
\end{equation}
is a positive self-adjoint operator in the Hilbert space of square
integrable functions of $r_\star$, and, therefore, all solutions of
the perturbative equations of motion with compact support initial
conditions are bounded.

However, as we will see, the effective
potentials may be non-positive definite for certain choices of
the parameters in Eqs.(\ref{cfm})-(\ref{zeroSch}).
\emph{Nevertheless, even when the effective potential is not
positive definite, we do not observe unbounded solutions.}

Using both high order WKB method and direct numerical integration
of the equations of motion a numerical search for quasinormal
modes with positive imaginary part was performed for scalar and
gravitational perturbations. One of the most important results in
this work is that \emph{no unstable mode was observed}.
Furthermore, the perturbative late-time tails have power-law
behavior (in one case an oscillatory decay with power-law envelope).

\begin{figure}[tp]
\resizebox{1\linewidth}{!}{\includegraphics*{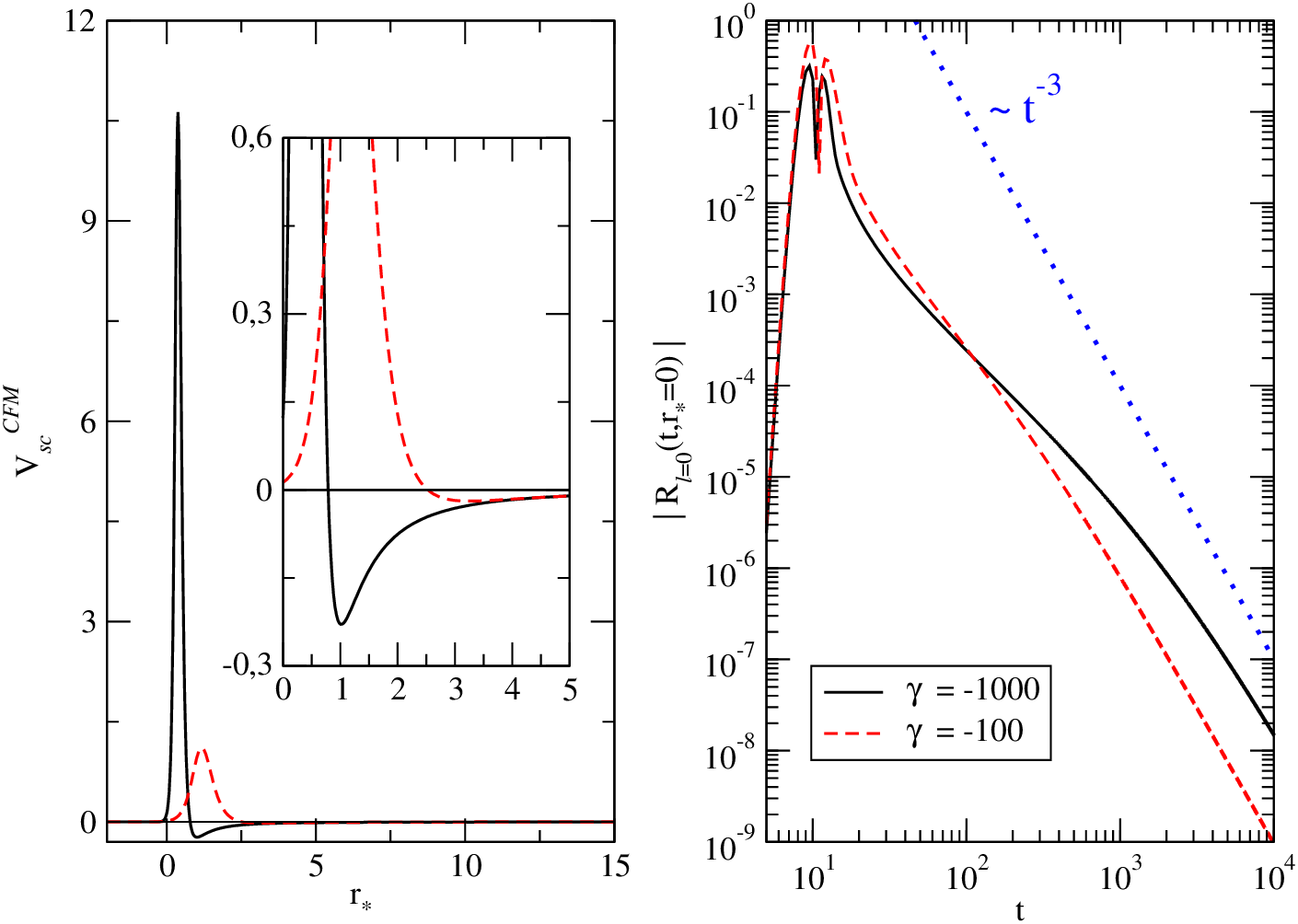}}
\caption{(Left) Effective potential for the scalar perturbations in
  the CFM background for very negative
  values of $\gamma$. Negative peaks are displayed in
  detail. (Right) Bounded evolution of the scalar field perturbation with such
  effective potentials. The dotted line is the late-time power-law
  tail. The parameters are $\ell=0$ and $M=1$.}
\label{highgamma1}
\end{figure}

\begin{figure}[tp]
\resizebox{1\linewidth}{!}{\includegraphics*{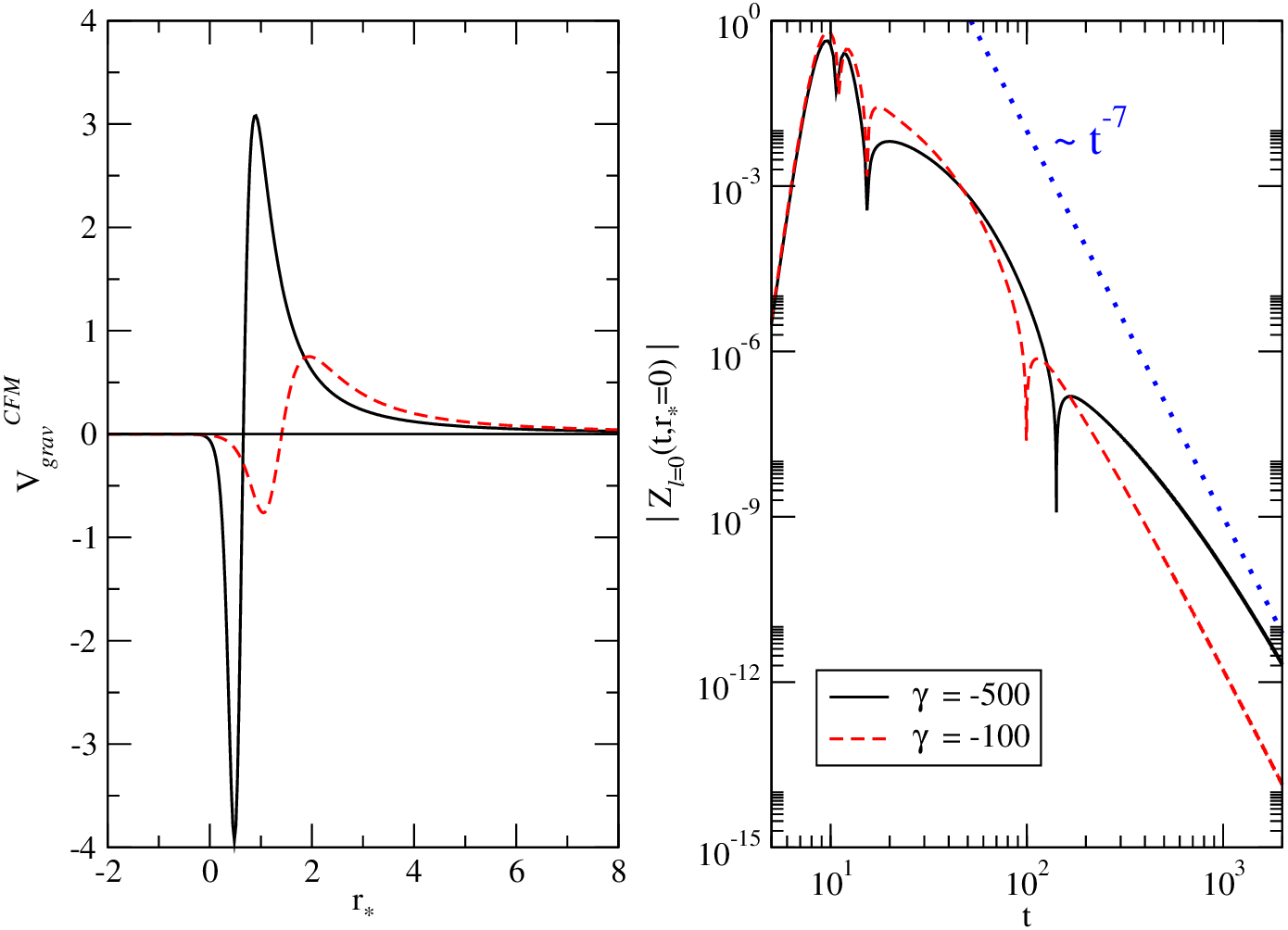}}
\caption{(Left) Effective potential for the axial gravitational
  perturbations in the CFM background for very negative
  values of $\gamma$. (Right) Bounded evolution of the gravitational
  field perturbation with such effective potentials. The dotted line
  is the late-time power-law tail. The parameters in the graphs are
  $\ell=2$ and $M=1$.}
\label{highgamma2}
\end{figure}

\begin{table*}[tp]
\label{mqn-sc-cfm}
\caption{Fundamental quasinormal frequencies for the scalar
  perturbation around the CFM black hole for several values of the
  parameters $\gamma$ and $\ell$. The black hole mass is set to $M=1$.}
\begin{ruledtabular}
\begin{tabular}{cccccccc}
\multicolumn{2}{l}{}                           &
\multicolumn{2}{c}{Direct Integration}         &
\multicolumn{2}{c}{WKB-$3^{th}$ order}                    &
\multicolumn{2}{c}{WKB-$6^{th}$ order}                \\ \\
$\ell$         & $\gamma$        & 
Re($\omega_0$) & -Im($\omega_0$) & 
Re($\omega_0$) & -Im($\omega_0$) & 
Re($\omega_0$) & -Im($\omega_0$) \\ \\
\hline \\
1 & -5  & 0.28580 & 0.19779 & 0.208204 & 0.225080 & 0.305499 & 0.181027 \\
1 & -2  & 0.29201 & 0.16608 & 0.276143 & 0.181776 & 0.309350 & 0.122589 \\
1 & 0   & 0.29337 & 0.14138 & 0.299359 & 0.154590 & 0.236213 & 0.165871 \\
1 & 1   & 0.29400 & 0.12853 & 0.298384 & 0.134468 & 0.252449 & 0.171879 \\
1 & 2   & 0.29415 & 0.11378 & 0.294679 & 0.115208 & 0.293168 & 0.120632 \\
1 & 3   & 0.29283 & 0.098045 & 0.291114 & 0.0980014 & 0.292910 & 0.0977616 \\
1 & 3.9 & 0.29076 & 0.082598 & 0.287181 & 0.0820285 & 0.289628 & 0.0811812 \\  \\
2 & -5  & 0.48053 & 0.18951 & 0.488726 & 0.207518 & 0.433498 & 0.183580 \\
2 & -2  & 0.48266 & 0.16069 & 0.488518 & 0.166137 & 0.451478 & 0.193079 \\
2 & 0   & 0.48413 & 0.13815 & 0.486195 & 0.139258 & 0.485925 & 0.142572 \\
2 & 1   & 0.48449 & 0.12570 & 0.485289 & 0.126043 & 0.485911 & 0.126758 \\
2 & 2   & 0.48447 & 0.11206 & 0.484420 & 0.112159 & 0.484691 & 0.112283 \\
2 & 3   & 0.48317 & 0.097097 & 0.483211 & 0.09680485 & 0.483642 & 0.0967661 \\
2 & 3.9 & 0.48178 & 0.080778 & 0.481091 & 0.08098300 & 0.481705 & 0.0808983 \\ \\
\end{tabular}
\end{ruledtabular}
\end{table*}

\begin{table*}[tp]
\label{mqn-ax-cfm}
\caption{Fundamental quasinormal frequencies for the axial gravitational
  perturbation around the CFM black hole for several values of the
  parameters $\gamma$ and $\ell$. The black hole mass is set to $M=1$.}
\begin{ruledtabular}
\begin{tabular}{cccccccc}
\multicolumn{2}{l}{}         &
\multicolumn{2}{c}{Direct Integration}         &
\multicolumn{2}{c}{WKB-$3^{th}$ order}         &
\multicolumn{2}{c}{WKB-$6^{th}$ order}     \\ 
$\ell$         & $\gamma$        & 
Re($\omega_0$) & -Im($\omega_0$) & 
Re($\omega_0$) & -Im($\omega_0$) & 
Re($\omega_0$) & -Im($\omega_0$) \\ \\
\hline \\
2 & -5  & 0.36409 & 0.18017 & 0.401345 & 0.197274 & 0.418575 & 0.193276 \\
2 & -2  & 0.37049 & 0.16062 & 0.391442 & 0.163223 & 0.402747 & 0.163222 \\
2 & 0   & 0.37359 & 0.13539 & 0.384611 & 0.137147 & 0.389781 & 0.139044 \\
2 & 1   & 0.37457 & 0.12138 & 0.381053 & 0.122624 & 0.383017 & 0.124656 \\
2 & 2   & 0.37483 & 0.10604 & 0.377306 & 0.106767 & 0.377126 & 0.107996 \\
2 & 3   & 0.37368 & 0.088957 & 0.373162 & 0.0892174 & 0.373619 & 0.0888910 \\
2 & 3.9 & 0.36961 & 0.072435 & 0.368552 & 0.0717786 & 0.371935 & 0.0712303 \\ \\
3 & -5  & 0.59476 & 0.18365 & 0.608026 & 0.191340 & 0.613069 & 0.194241  \\
3 & -2  & 0.59835 & 0.15845 & 0.604567 & 0.159482 & 0.605032 & 0.161682  \\
3 & 0   & 0.59993 & 0.13527 & 0.602594 & 0.135600 & 0.601901 & 0.136695  \\
3 & 1   & 0.60033 & 0.12238 & 0.601646 & 0.122520 & 0.601033 & 0.123070  \\
3 & 2   & 0.60026 & 0.10832 & 0.600614 & 0.108374 & 0.600375 & 0.108525  \\
3 & 3   & 0.59947 & 0.092690 & 0.599265 & 0.0927284 & 0.599443 & 0.0927025  \\
3 & 3.9 & 0.59700 & 0.077176 & 0.597227 & 0.0767434 & 0.597584 & 0.0767411  \\
\end{tabular}
\end{ruledtabular}
\end{table*}

\subsection{CFM brane black holes}

We first consider scalar perturbations in the CFM scenario. The
tortoise coordinate $r_\star$ after the explicit integration is
\begin{equation}
r_{\star}(r) =   T_{1}(r) +  T_{2}(r) +  T_{3}(r)
\end{equation}
with
\begin{equation}
T_{1}(r) = \frac{1}{2} \sqrt{(2r-\gamma M)(2r-3M)} \, , \\
\end{equation}
\begin{equation}
T_{2}(r) =
\frac{M(5 + \gamma)}{4} \ln \left[4r - M(3+\gamma) + 2 T_1(r) \right]
, \\
\end{equation}
\begin{gather}
T_{3}(r) = -\frac{2M}{\sqrt{4-\gamma}} \nonumber \\
\times \ln \left[\frac{M(5 - \gamma)r -
    M^{2}(6-\gamma) + M\sqrt{4-\gamma} \, T_{1}(r) }{r-2M}\right] \, .
\end{gather}

The scalar and axial gravitational effective potentials for
perturbations in the CFM background (respectively, 
$V_{sc}^{CFM}$ and $V_{grav}^{CFM}$) in terms of the parameters $M$ and
$\gamma$ are given by
\begin{gather}
V_{sc}^{CFM}(r)  =  \left(1 - \frac{2M}{r}\right) \left[ \frac{\ell
    (\ell + 1)}{r^2} + \frac{2M}{r^3} \right. \nonumber \\
+ \left. \frac{ M (\gamma - 3) (r^2 - 6Mr + 6M^2)}{r^3(2r - 3M)^2}
\right] \, 
\label{pot_escalar_cfm}
\end{gather}
and
\begin{gather}
 V_{grav}^{CFM}(r) = \left( 1 - \frac{2M}{r} \right) \left[ \frac{\ell (\ell +
 1)}{r^{2}} - \frac{6M}{r^{3}} \nonumber  \right.\\
- \left. \frac{M (\gamma -3) (5r^{2} - 20 M r + 18M^{2}) }{r^{3} (2r -
  3M)^{2}}   \right] \, .
\label{pot_axial_cfm}
\end{gather}
Setting $\gamma = 3$ we recover the usual expressions for
perturbations around the four dimensional Schwarzschild black hole.

A basic feature of the effective potentials $ V_{sc}^{CFM}$ and $
V_{grav}^{CFM}$ is that they are not positive definite. Typically,
for negative enough values of the parameter $\gamma$ (with large
$|\gamma|$) a negative peak in the effective potential will show
up. It is no longer obvious that the scalar and gravitational
perturbations will be stable. 

If the effective potential is not positive definite (and cannot be
approximated by a positive definite one), the WKB semi-analytical
formulae will usually not be applicable, but direct integration
techniques will. Using the later approach an extensive search for
unstable solutions was made. One important result in this work is
that even for very high values of $\gamma$, \emph{the (scalar and
gravitational) perturbative dynamics is always stable}. This is
illustrated in Figs. \ref{highgamma1} and \ref{highgamma2}, where 
we show non-positive definite effective potentials and the
corresponding (bounded) field evolution.

The overall picture of the perturbative dynamics for the effective
potentials hereby considered is the usual one. After a brief transient
regime, the quasinormal mode dominated phase follows, and, finally,
at late times a power-law tail dominates. 

For the fundamental multipole mode ($\ell=0$) the effective potential
will not be positive definite for any value of the parameter $\gamma$. 
Direct integration shows that the field evolution is
always bounded for a great range of variation of $\gamma$. This point
is illustrated in Figs. \ref{highgamma1} and 2, where
we have selected rather large values of $\gamma$. Indeed, it is
observed that the decay is dominated from very early time by the
power-law tail. Therefore, it is very difficult to estimate the
quasinormal frequencies directly from this ``time--dependent''
approach. The WKB-type expressions are not applicable if $\ell=0$ for
two reasons: the effective potentials are not positive for $r$ larger
than a certain value, and it is well known that this method works
better with $\ell < n$, where $n$ is the overtone number. 

With small but non-zero values of $\ell$ the quasinormal frequencies
can be accurately estimated. As it is shown in Tables I and
II, the concordance with the WKB results is reasonable,
except for some values of $\gamma$ (typically around $\gamma=0$).   

For large values of $\ell$ an analytical expression for the
quasinormal frequencies can be obtained. Expanding the effective
potential in terms of small values of $1/\ell$ and using the WKB
method in the lowest order (which is exact in this limit) we find 
\begin{equation}
\textrm{Re} (\omega_{n}) = \frac{\ell}{3\sqrt{3} M} \, ,
\end{equation}
\begin{equation}
\textrm{Im} (\omega_{n}) = \sqrt{\frac{2\ell}{3 M^{2}} \left(n +
  \frac{1}{2}  \right) } \, .
\end{equation}

\begin{figure}[tp]
\resizebox{1\linewidth}{!}{\includegraphics*{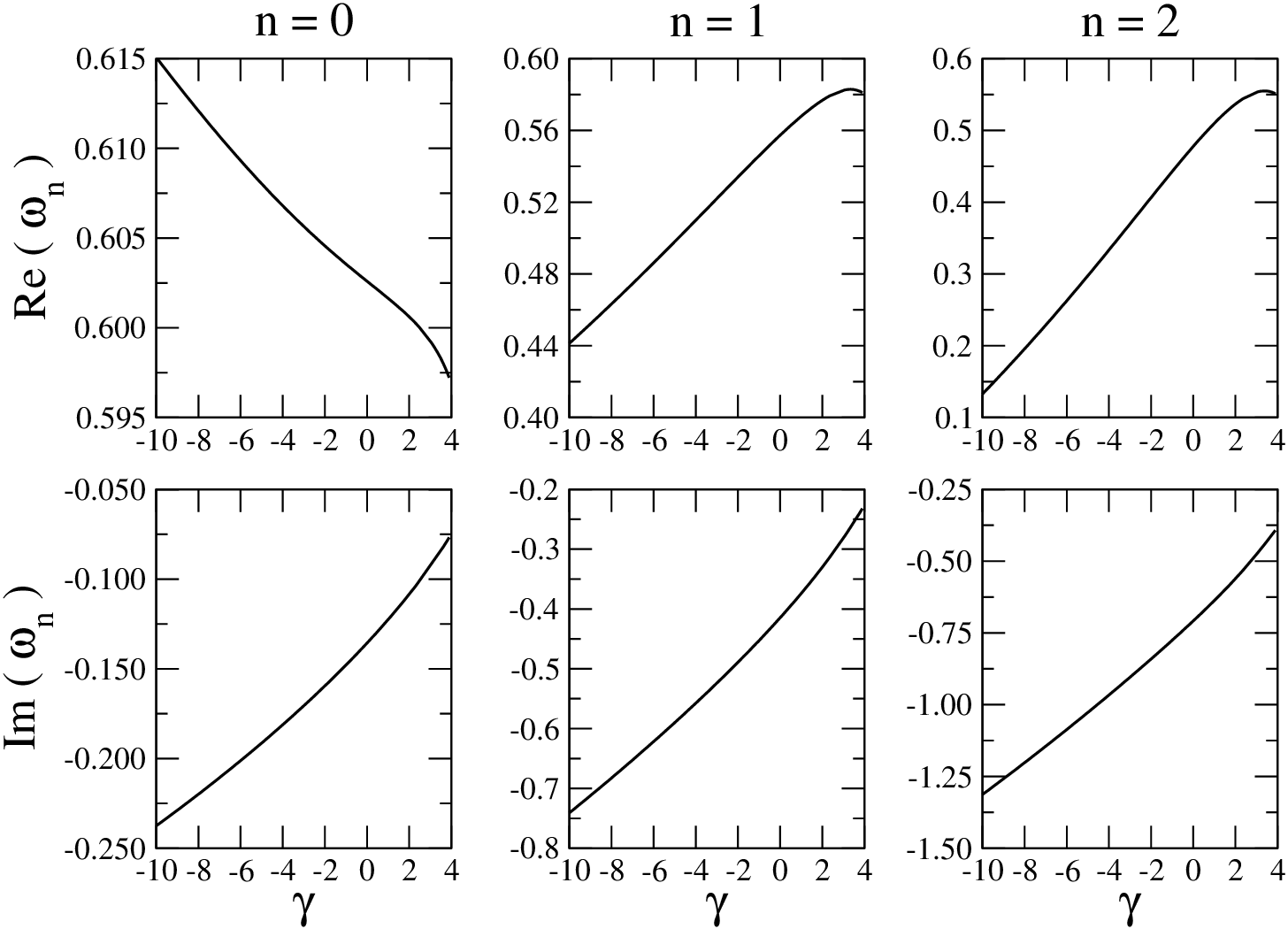}}
\caption{Dependence of gravitational perturbation quasinormal
  frequencies on $\gamma$ in the CFM geometry. The results are
  qualitatively similar for the scalar perturbation. The
  parameters are $\ell=3$ and $M=1$.}
\label{mqn_ax}
\end{figure}

As it can be seen from the data in tables
I and II and Fig. \ref{mqn_ax}, the
dependence of the frequencies with the parameter $\gamma$ is very
weak, although not trivial. 
In a large variation range of
$\gamma$ the absolute value of $\textrm{Im} (\omega_0)$ is a
monotonically decreasing function, while  $\textrm{Re} (\omega_0)$
typically has maximum points.

The late-time behavior of the perturbations considered here can be
treated analytically. Far from the black hole the scalar
effective potential in terms of $r_{\star}$ assumes the form,
\begin{equation}
V_{sc}^{CFM}(r_\star) \sim
            \begin{cases}
            \frac{2M}{r_{\star}^{3}}   & \textrm{with} \,\, \ell=0 \, \\
            \frac{\ell(\ell + 1)}{r_{\star}^{2}}
            + \frac{4M \ell (\ell + 1) \ln
            (r_{\star}/2M)}{r_{\star}^{3}} &  \textrm{with} \,\,
            \ell>0 \, .
            \end{cases} 
\end{equation}
It is then shown \cite{Price,Ching} that with the initial data
having compact support a potential with this form has a late-time
tail
\begin{equation}
R_{\ell}^{CFM} \sim t^{-(2\ell + 3)} \, .
\end{equation}
Therefore, at asymptotically late times \emph{the perturbation
decays as a power-law tail for any value of the parameter
$\gamma$}. This is a strong indication that the models are indeed
stable. This point is illustrated in Figs. \ref{highgamma1} and
\ref{highgamma2}. It is reminiscent from a similar behavior of the
Gauss Bonnet term added to Einstein gravity in higher dimensions,
which was recently treated in \cite{abdkonomol}. Although the
result is formally valid also for the gravitational perturbations we have
considered, it should be noted that in the simplified model
developed in this paper the back-reaction from the bulk, which can
modify the tail presented here, was neglected.

\begin{table*}[tp]
\label{mqnsczm}
\caption{Fundamental quasinormal frequencies for the scalar
  perturbation around the ``zero mass'' black hole for several values
  of the $C$ and $\ell$. The parameter $h$ is set to $h=1$.} 
\begin{ruledtabular}
\begin{tabular}{cccccccc}
\multicolumn{2}{l}{}                           &
\multicolumn{2}{c}{Direct Integration}         &
\multicolumn{2}{c}{WKB-$3^{th}$ order}         &
\multicolumn{2}{c}{WKB-$6^{th}$ order}        \\ \\
$\ell$         & $C$        &
Re($\omega_0$) & -Im($\omega_0$) &
Re($\omega_0$) & -Im($\omega_0$) &
Re($\omega_0$) & -Im($\omega_0$) \\ \\
\hline \\
1 & 0   &    -    &    -    & 0.728358 & 0.232800 & 0.746504 & 0.230697 \\
1 & 0.1 & 0.74494 & 0.24551 & 0.730716 & 0.247607 & 0.749727 & 0.245155 \\
1 & 0.5 & 0.75196 & 0.30135 & 0.732795 & 0.303801 & 0.755273 & 0.301191 \\
1 & 0.9 & 0.75437 & 0.35023 & 0.726005 & 0.356124 & 0.751242 & 0.357060 \\
1 & 1.0 & 0.75410 & 0.36201 & 0.722899 & 0.368853 & 0.748461 & 0.371463 \\
1 & 1.1 & 0.75354 & 0.37352 & 0.719182 & 0.381471 & 0.745143 & 0.385894 \\
1 & 2.0 & 0.74548 & 0.46711 & 0.660519 & 0.494129 & 0.732352 & 0.479133 \\ \\ 
2 & 0   &    -     &    -   & 1.242071 & 0.230667 & 1.246220 & 0.231061 \\
2 & 0.1 & 1.24534 & 0.21001 & 1.243863 & 0.245695 & 1.248248 & 0.246014 \\
2 & 0.5 & 1.25534 & 0.28301 & 1.246290 & 0.299897 & 1.251377 & 0.300033 \\
2 & 0.9 & 1.25072 & 0.34681 & 1.244715 & 0.347044 & 1.250452 & 0.347062 \\
2 & 1.0 & 1.24989 & 0.35677 & 1.243914 & 0.358055 & 1.249839 & 0.358066 \\
2 & 1.1 & 1.25377 & 0.36031 & 1.242968 & 0.368816 & 1.249090 & 0.368835 \\
2 & 2.0 & 1.24523 & 0.44843 & 1.227403 & 0.456844 & 1.235779 & 0.459396 \\  \\
\end{tabular}
\end{ruledtabular}
\end{table*}

\subsection{``Zero mass'' brane black holes}

\begin{figure}[tp]
\resizebox{1\linewidth}{!}{\includegraphics*{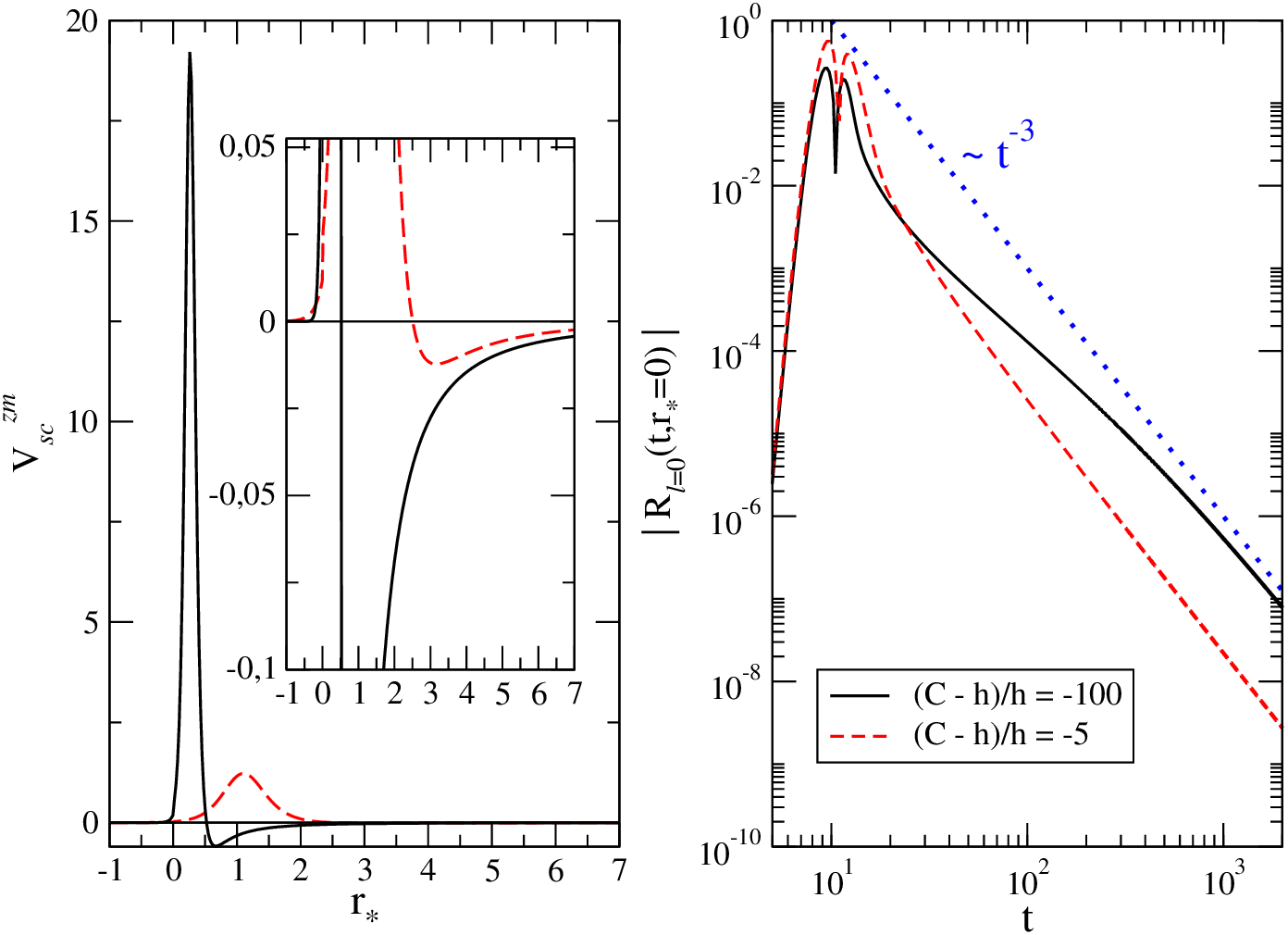}}
\caption{(Left) Effective potential for the scalar perturbations in
  the ``zero mass'' black hole background with high  values of
  $C$. Negative peaks are displayed in detail. (Right) Bounded
  evolution of the scalar field perturbation with such effective
  potentials. The dotted line is the late-time power-law tail. The
  parameters are $\ell=0$ and $h=1$.} 
\label{highC1}
\end{figure}

The treatment in section IV is general enough to include also the
case of perturbations around the ``zero mass'' brane black hole.
Using the metric (\ref{zeroSch}) the scalar and axial
gravitational perturbations are described by wave equations
similar to Eq.(\ref{wave_equation1}) with \linebreak effective potentials
given by
\begin{gather}
V_{sc}^{zm}(r) =  \left(1 - \frac{h^{2}}{r^{2}}\right) \left[ \frac{\ell
    (\ell + 1)}{r^2} + \frac{2 h^{2}}{r^4} \right. \nonumber \\
- \left. \frac{C - h}{(2r^{2} - h^{2})^{3/2} } 
\left( 1 - \frac{5 h^{2}}{r^{2}} + \frac{2 h^{4}}{r^{4}} \right)
  \right ] \, ,
\label{pot_escalar_zm}
\end{gather}
and
\begin{gather}
 V_{grav}^{zm}(r) = \left(1 - \frac{h^{2}}{r^{2}}\right)
 \left[ \frac{\ell (\ell + 1)}{r^2} - \frac{4 h^{2}}{r^{4}}
  \right. \nonumber\\ 
+ \left. \frac{C - h}{(2r^{2} -
 h^{2})^{3/2}} \left(5 - \frac{11 h^{2}}{r^{2}}  + \frac{4 h^4}{r^4}
 \right) \right]  \, .
\label{pot_axial_zm}
\end{gather}
Again, the effective potentials can be non-positive definite for
specific choices of parameters, as illustrated in Fig. \ref{highC1}
(left). For example, if $\ell=0$ and $C>h$,  $V_{sc}^{zm}$ will
 not  be positive definite. If  $\ell>0$, $V_{sc}^{zm}$ and
$V_{grav}^{zm}$ will be non-positive definite for high enough values
of $C$.    

\begin{figure}[tp]
\resizebox{1\linewidth}{!}{\includegraphics*{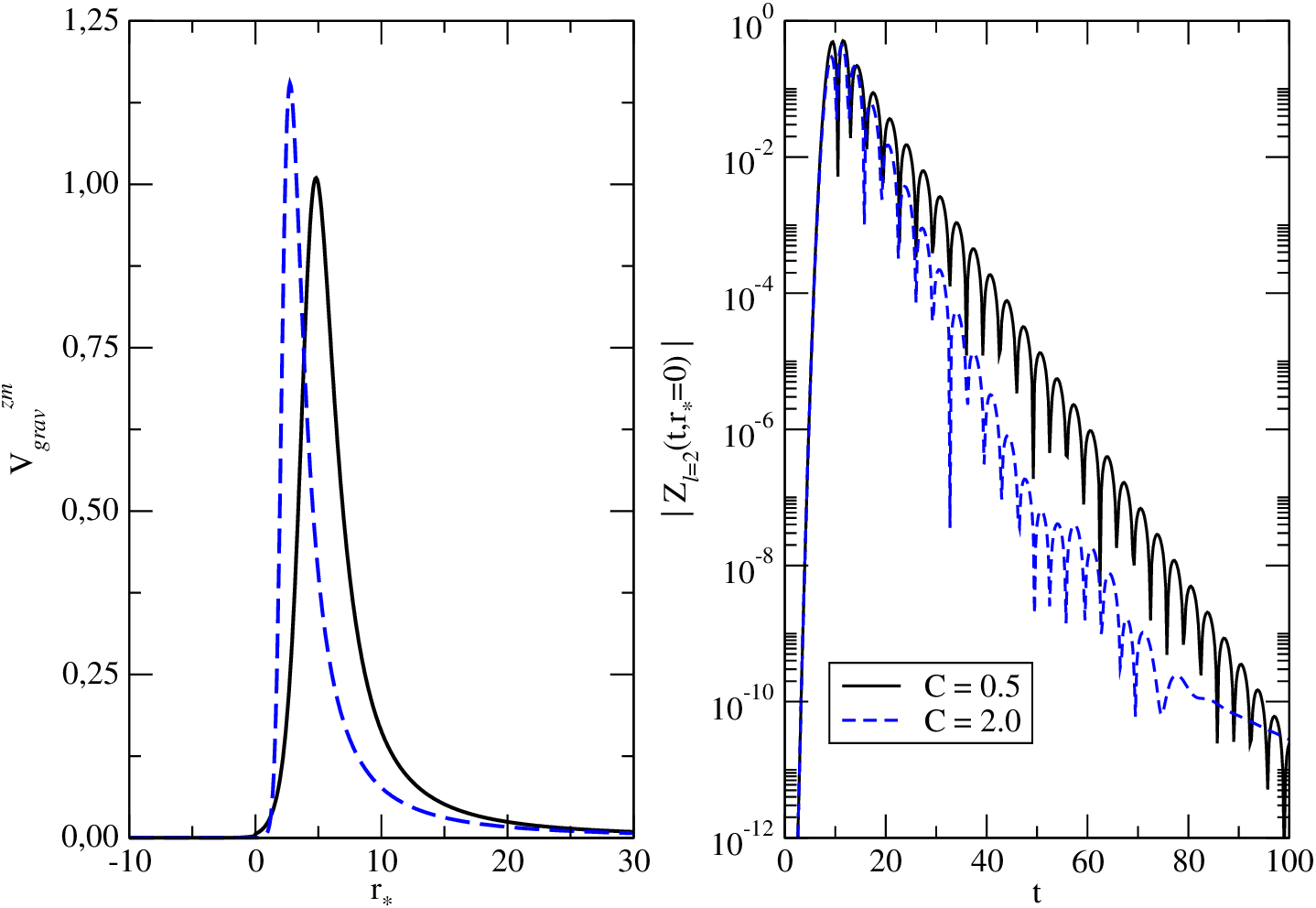}}
\caption{(Left) Typical effective potential for the gravitational perturbations in
  the ``zero mass'' black hole background. (Right) Bounded
  evolution of the gravitational field perturbation with such effective
  potentials. The parameters are $\ell=2$ and $h=1$.} 
\label{zm-ax}
\end{figure}

Except for $C = h$,  an explicit expression for the tortoise
coordinate was not found. Nevertheless, the numerical integration is
possible. The semi-analytical WKB approach was also used to compute quasinormal
frequencies.  The concordance is excellent. The WKB formulas seem to
be more reliable in the present case. With the choice $C=1$ we recover
some results considered in \cite{roman}. Again, an extensive search
for unstable modes was performed. Some calculated frequencies are
shown in Tables III and IV. Our results show that \emph{the dynamics
of the scalar and axial gravitational perturbations is always stable
in the ``zero mass'' background}. We illustrate this point in
Figs. \ref{highC1} and \ref{zm-ax}.

Analytical expressions for the quasinormal frequencies for the scalar and
gravitational perturbations can be obtained in the
limit of large multipole index $\ell$. As done in the CFM geometry, we
obtain
\begin{equation}
\textrm{Re} (\omega_{n}) = \frac{\ell}{2 h} \, ,
\end{equation}
\begin{equation}
\textrm{Im} (\omega_{n}) =  \sqrt{\frac{\ell}{\sqrt{2} h} \left(n +
  \frac{1}{2}  \right)}  \, . 
\end{equation}

\begin{figure}[tp]
\resizebox{1\linewidth}{!}{\includegraphics*{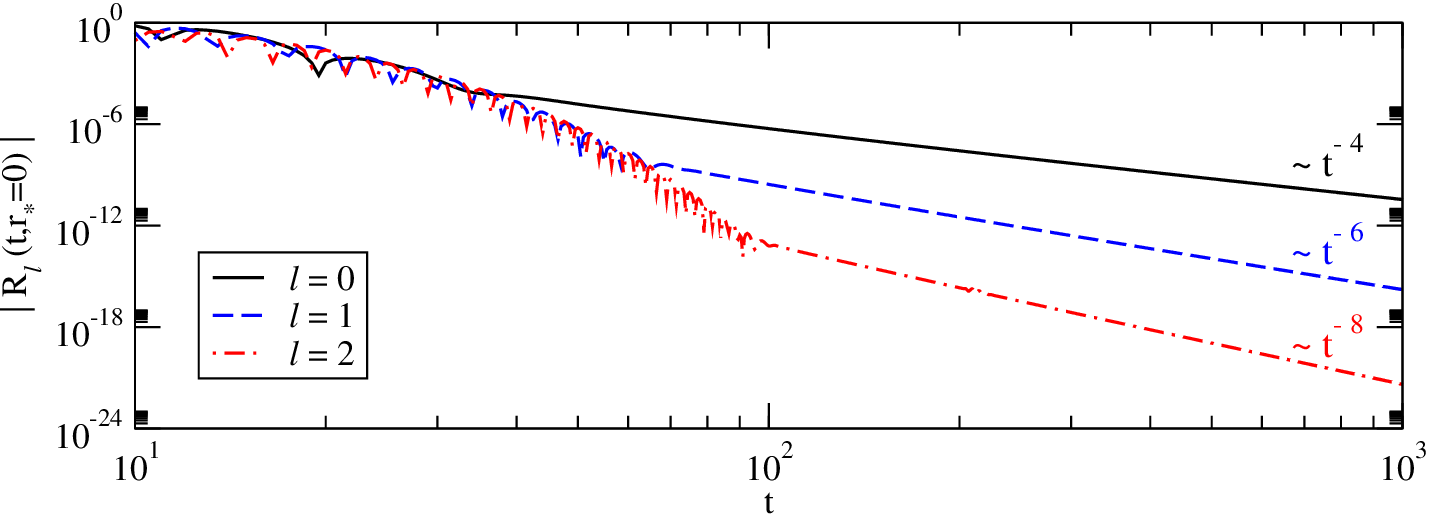}}
\caption{Bounded evolution of the scalar field perturbation in the
  ``zero mass'' black hole background with $C=h$ for several values
  of $\ell$. After the quasinormal mode phase a
  power-law tail is observed. 
The tail dependence with $\ell$ obeys Eq. (\ref{tail_Ceqh}). 
In the graphs the parameter $h$ was set to $h=1$.} 
\label{zm_heqC}
\end{figure}

We expect that the late-time behavior of the gravitational
perturbations should be dominated by back scattering from the bulk,
not considered here. But the tail contribution to the scalar decay can
be analytically treated, at least in the limit where $r \gg h$. In
this case if $C>h$ or $0<C<h$, the effective potential $V_{sc}^{zm}$
is approximated by  
\begin{equation}
V_{sc}^{zm}(r_\star) \sim
            \begin{cases}
            \frac{C - h}{\sqrt{2} r_{\star}^{3}}  &  \textrm{with}
            \,\, \ell=0 \, , \\
            \frac{\ell(\ell + 1)}{r_{\star}^{2}}
            + \frac{2 (C - h) \ell (\ell + 1) \ln
            (r_{\star}/h)}{\sqrt{2} r_{\star}^{3}} &  \textrm{with}
            \,\, \ell>0 \, .
            \end{cases}
\end{equation}
Again, we observe that (\cite{Price,Ching})  with the initial data
having compact support the tail has the form
\begin{equation}
R_{\ell}^{zm} \sim t^{-(2\ell + 3)} \,\,\, \textrm{with} 
\,\,\, 0<C<h   \,\,\,\textrm{or}  \,\,\,  C>h  \, .
\end{equation}

An interesting limit is when $C=h$. In this case the explicit
expression for the tortoise coordinate according to the usual
definition in Eq. (\ref{rstar}) is 
\begin{equation}
r_{\star}(r) = r + \frac{h}{2} \ln \left( \frac{r}{h} - 1 \right)
- \frac{h}{2} \ln \left( \frac{r}{h} + 1 \right) \, .
\end{equation}
The effective potential $V_{sc}^{zm}$ with $C=h$ is approximated by
\begin{equation}
V_{sc}^{zm}(r_\star) \sim
            \begin{cases}
            \frac{2 h^{2}}{r_{\star}^{4}}  & \textrm{with}  \,  \, \ell=0 \, , \\
            \frac{\ell(\ell + 1)}{r_{\star}^{2}}
            - \frac{2\ell(\ell + 1) h^{2}}{r_{\star}^{4}} &
            \textrm{with}  \,  \, \ell>0 \, .
            \end{cases}
\end{equation}
In this limit a power-law still dominates the late-time decay. But
its dependence with the multipole index $\ell$ is different,
\begin{equation}
R_{\ell}^{zm} \sim t^{-(2\ell + 4)} \,\,\, \textrm{with} 
\,\,\, C=h  \, .
\label{tail_Ceqh}
\end{equation}
This point is illustrated in Fig. \ref{zm_heqC}. 

As observed in the CFM model, for the non-extreme ``zero mass'' model
\emph{the scalar perturbation decays as a power-law tail} suggesting that
the model is stable.  

\begin{figure}
\resizebox{1\linewidth}{!}{\includegraphics*{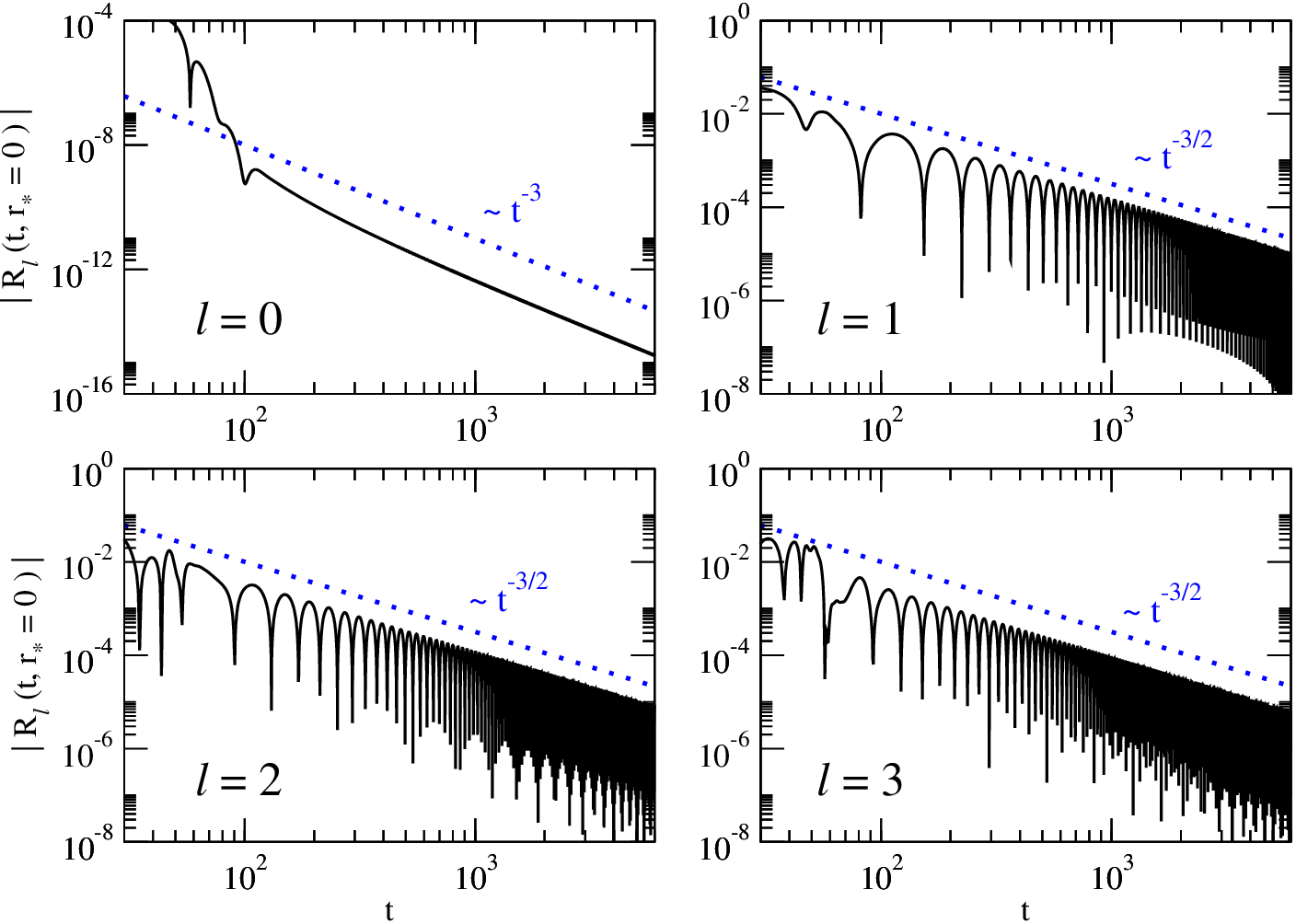}}
\caption{Bounded evolution of the scalar field perturbation in the
  extreme ``zero mass'' black hole background ($C=0$), for several values
  of $\ell$. If $\ell=0$, the decay is dominated by a power-law tail
  ($t^{-3}$). If  $\ell>0$,  the decay is dominated by an oscillatory
  tail, whose envelope is $t^{-3/2}$. In the graphs the parameter $h$
  was set to $h=1$.}  
\label{zm_extrem}
\end{figure}

The qualitative picture of the field evolution in the ``zero mass''
black hole --- quasinormal mode followed by power-law tail --- changes
drastically when the extreme case ($C=0$) is considered (see
Fig. \ref{zm_extrem} ). If $\ell=0$, we observe the usual power-law
tail dominating the late-time decay. But when $\ell>0$, the simple
power-law tail is replaced by an oscillatory decay with a power-law
envelope, 
\begin{equation}
R_{\ell}^{zm} \sim t^{-3/2} \sin \left( \omega_{\ell} \times t \right) 
\,\,\, \textrm{with} \,\,\, C=0 \,\, \textrm{and}  \,\, \ell>0  \, .
\label{tail_extreme}
\end{equation}

\begin{figure}
\resizebox{1\linewidth}{!}{\includegraphics*{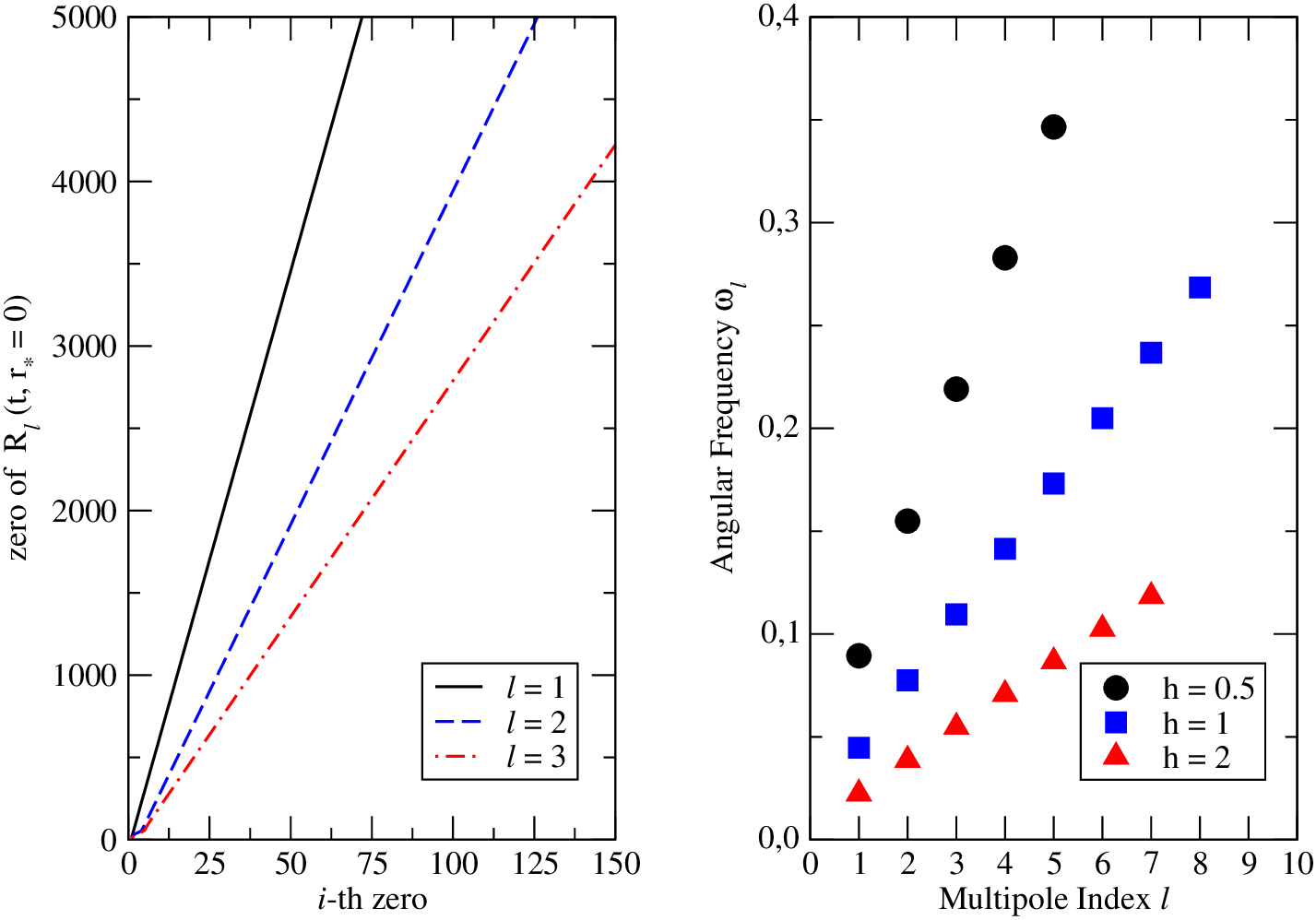}}
\caption{ (Left) Numerical value of $t$ where the scalar wave function  is
  zero in the extreme ``zero mass'' black hole. Straight lines imply
  that the period of oscillation is a constant. The parameter $h$ is set
  to $h=1$. (Right) Dependence of
  the angular frequency $\omega_{\ell}$ in Eq.(\ref{tail_extreme})
  with $\ell$ for several values of $h$.}   
\label{zm_extrem2}
\end{figure}

Therefore, for $\ell>0$ the power index ($-3/2$) is independent of  the multipole
index $\ell$.  The angular frequency $\omega_{\ell}$
for large times approaches a constant, as we can see in
Fig. \ref{zm_extrem2} (left). The angular frequency is well approximated
by a linear function of $\ell$, as indicated in Table V
for some values of $h$.  This result implies that the dominating
contributions in the late-time decay are the modes with $\ell>0$,
\textit{i.e.}, the power-law enveloped oscillatory terms. We  also
observe that these tails dominate from very early times, so that it
was not possible to estimate the quasinormal frequencies in the
extreme case (as indicated in the first lines of Tables III and IV).

\begin{table*}[tp]
\label{mqngravzm}
\caption{Fundamental quasinormal frequencies for the axial gravitational
  perturbation around the ``zero mass'' black hole for several values
  of the $C$ and $\ell$. The parameter $h$ is set to $h=1$.} 
\begin{ruledtabular}
\begin{tabular}{cccccccc}
\multicolumn{2}{l}{}                           &
\multicolumn{2}{c}{Direct Integration}         &
\multicolumn{2}{c}{WKB-$3^{th}$ order}                       &
\multicolumn{2}{c}{WKB-$6^{th}$ order}                        \\ \\
$\ell$         & $C$        &
Re($\omega_0$) & -Im($\omega_0$) &
Re($\omega_0$) & -Im($\omega_0$) &
Re($\omega_0$) & -Im($\omega_0$) \\ \\
\hline \\
2 & 0   &   -     &    -    & 0.934530 & 0.191198 & 0.934386 & 0.219631 \\
2 & 0.1 & 0.94412 & 0.21924 & 0.935381 & 0.205749 & 0.964066 & 0.215448 \\
2 & 0.5 & 0.94783 & 0.26533 & 0.938066 & 0.264005 & 0.958654 & 0.249590 \\
2 & 0.9 & 0.94334 & 0.30949 & 0.942035 & 0.318153 & 0.929971 & 0.317851 \\
2 & 1.0 & 0.94207 & 0.31993 & 0.943255 & 0.330875 & 0.928449 & 0.334460 \\
2 & 1.1 & 0.94069 & 0.33016 & 0.944551 & 0.343278 & 0.928937 & 0.350243 \\
2 & 2.0 & 0.92442 & 0.41343 & 0.958486 & 0.442675 & 0.973998 & 0.456121 \\ \\
3 & 0   &    -    &    -    & 1.537001 & 0.214671 & 1.538619 & 0.215204 \\
3 & 0.1 & 1.50122 & 0.25435 & 1.537512 & 0.229197 & 1.539283 & 0.229786 \\
3 & 0.5 & 1.53958 & 0.27762 & 1.536138 & 0.280829 & 1.538697 & 0.280900 \\
3 & 0.9 & 1.53905 & 0.32341 & 1.532735 & 0.325500 & 1.534597 & 0.324529 \\
3 & 1.0 & 1.53757 & 0.33402 & 1.531809 & 0.335949 & 1.533070 & 0.334820 \\
3 & 1.1 & 1.53594 & 0.34411 & 1.530883 & 0.346171 & 1.531369 & 0.344973 \\
3 & 2.0 & 1.52161 & 0.42100 & 1.523732 & 0.430806 & 1.513596 & 0.433952 \\ \\
\end{tabular}
\end{ruledtabular}
\end{table*}

\begin{table}[tp]
\label{omega_ell}
\caption{Oscillatory frequency of the tail in the extreme ``zero
  mass'' ($C=0$) black hole for several values of $h$.} 
\begin{ruledtabular}
\begin{tabular}{cc}
$h$         & Angular Frequency $\omega_{\ell}$ ($\ell>0$) \\
\hline \\
0.5 & $0.02593  + 0.06421 \times \ell$ \\
1.0 & $0.01347  + 0.03191 \times \ell$ \\
2.0 & $0.006683 + 0.01597 \times \ell$ 
\end{tabular}
\end{ruledtabular}
\end{table}

\section{Conclusions}

In this work we studied brane black holes from the point of view of a
brane observer. We considered the two family solutions found by
Bronnikov \textit{et al.} \cite{bron} in order to derive the Bekenstein
entropy bound and the one-loop correction to the Bekenstein-Hawking
formula using the 't Hooft brickwall method. In addition, we performed
scalar and axial gravitational perturbations in two specific examples
of these families. With these perturbations we were able to analyze 
the dynamics and stability of the black hole solutions.

The results of the black hole thermodynamics study show that the
entropy bound continues to be independent of the black hole
parameters. Thus, the presence of the bulk does not affect the
universality of the entropy bound for a brane observer, as ourselves,
reinforcing the Generalized Second Law. Moreover, applying the 't
Hooft's brickwall method to both black hole families we see that the
entropy correction takes the same form as that of a Schwarzschild
black hole when written in terms of its own black hole
parameters. Therefore, as the correction is linearly dependent on the
area, it can be absorbed in a renormalized gravitational constant.

One of the most important results in this paper came from the
perturbative dynamics. We should stress that the assumption $\delta
E_{\alpha\beta} =0$ was necessary in order to solve the gravitational
perturbation equation (\ref{gravperteq}) without any knowledge of the
bulk structure. This vanishing tidal effect is perfectly justified
when the perturbation energy is lower than the threshold of the
Kaluza-Klein massive modes. Likewise, as we neglect the bulk
back-reaction, our analysis does not describe the perturbation
late-time behavior. Our results show no unstable mode in the scalar
and gravitational analysis. In addition, the late-time tails display a 
power-law behavior what enforces their stability.

In the case of CFM black hole even if the
effective potential is not positive--definite the quasinormal
modes are stable (negative imaginary part). The agreement of the several
methods employed in the calculation is good for  $\ell$ not
too small.

On the other hand, in the case of the ``zero mass'' black hole 
we observe a richer picture. The scalar and gravitational field evolution is always
bounded suggesting that this class of models is stable. But the late
time decay of the matter field strongly depends on the parameters $C$
and $h$. If $C$ is non-zero and not equal to $h$, the late-time decay
is dominated by a power-law tail with the usual dependence on the
multipole parameter $\ell$. But if $C=h$, this dependence changes.
Finally, in the extreme regime ($C=0$) the late-time decay is
dominated by oscillatory modes with a power-law envelope. This
power index seems to be universal, not depending on $\ell$.

Summarizing, the thermodynamics in the class of models we considered is
consistent, while the dynamics  in specific backgrounds is stable in
the approach employed in the present work. \linebreak
While our results suggest
that the brane models presented are viable, the final check would be
the analysis of the continuation in the bulk of the geometries
presented here.

\begin{acknowledgments}
 This work was partially supported by Fundaç\~ao de Amparo \`a
 Pesquisa do Estado de S\~ao Paulo (FAPESP) and Conselho Nacional de
 Desenvolvimento Cient\'{\i}fico e Tecnol\'ogico (CNPq).
\end{acknowledgments}

\end{document}